# Polyhalogenated Molecules in the Polarizable Ellipsoidal Force Field Model

**Fang Liu**[*]

School of Information Engineering, Binzhou University, Binzhou, Shandong, China

liufang_ibi@webmail.hzau.edu.cn

liuf_bzu@163.com

Abstract — Polyhalogenated compounds are common in industrial, agricultural, and environmental applications. These compounds contain more halogen bonds than monohalogenated compounds. The presence of coupled σ-holes in the covalent halogen atoms, which demonstrates strong interplay in halogen bonds, should be carefully evaluated in force field optimization. In this study, a polarizable ellipsoidal force field model was successfully applied to many possible halogenated benzenes. The symmetry of the fitted parameters was reproduced without any additional restrictions. The optimized parameters for the anisotropic electrostatic potential showed good accuracy, stability, and transferability with reasonable physical meanings. The parameter fitting protocol was efficiently performed on a laptop, showing the potential of being completely parallelized for thousands of halogenated homologues.

## 1. Introduction

Halogenated molecules containing more than one covalent halogen atom form polyhalogenated compounds. Polyhalogenated molecules are widely used in industrial, agricultural, and environmental applications. For example, polybrominated diphenyl ethers and polychlorinated biphenyls are endocrine-disrupting compounds that induce a wide range of developmental and neurodegenerative effects.[1,2] Every covalent halogen atom in a polyhalogenated molecule may interact with other molecules via halogen bonds. The formation of halogen bonds between covalent halogen atoms with other molecules, as well as the coexistence of halogen bonds, especially in ortho- and meta-substituted compounds, play crucial roles in many chemical and biochemical processes.[3,4].

Nucleophilic attraction usually contributes to the formation of a positive center along the R–X direction, referred to as the sigma (σ)-hole.[5] The presence of a σ-hole around the covalent halogen atom is the prerequisite for the formation of a halogen bond. The σ-hole interactions cause the electrostatic potential of the covalent halogen atom to exhibit anisotropic charge distributions. In addition, quantitative molecular orbital analysis has proven that the excellent stability demonstrated by covalent compounds are attributed to the highest occupied molecular orbital (HOMO)–lowest unoccupied molecular orbital (LUMO) interactions between the halogen bond acceptor and donor.[6]

These bond features should be carefully considered in force field analysis, or at least be implicitly incorporated in force field parameters. However, the incorporation of these features for an atomic point-charge model in a traditional force field is challenging. In recent years, several force field models have been developed for halogen atoms or atoms with properties similar to those of halogen atoms, such as the positive extra-point charge or explicit σ-hole model,[7,8] force field for biological halogen bonds,[9] and polarizable ellipsoidal force field (PEff) model.[10]. The anisotropic electrostatic potential distribution in these models is generally recovered by providing additional force field parameters.

Previous studies have mostly focused on monohalogenated systems,[11-14] such as rational drug design[15,16] and protein-ligand complexes.[13,17] Simulation studies for polyhalogenated systems have also been reported. However, studies on the non-additive-coupled σ-hole interactions in polyhalogenated molecules are still limited. For instance, the PEff model proposed by Du et al. uses four parameters to describe the electrostatic potential around a covalent halogen atom. As the number of covalent halogen atoms increases, the degree of freedom to be fitted in the polyhalogenated molecules systematically increases, causing the parameter-fitting procedure to become intractable. This challenge has been identified in numerous many-objective optimization problems (MaOPs) in the field of mathematics.[18-20] In this regard, genetic algorithms[21,22] and machine-learning methods[23-25] have been identified as possible global optimization approaches. Taking the 1,2,3-tribromobenzene molecule as an example, the three covalent bromine atoms in the molecule are identified as three distinct optimization objects. Although the 12 parameters of the three covalent atoms can be directly optimized, the optimized parameters are highly dependent on the initial guess or fitting algorithms. In addition, although the parameters for the three bromine atoms can be fitted individually, the optimized parameters are strongly dependent on the order of the fitting procedure. As such, the symmetry of polyhalogenated molecules is usually destroyed without manual constraints, and the optimized parameters cannot provide reasonable physical meanings.

In the present study, the PEff model was applied to a series of polyhalogenated benzene molecules. The parameter-fitting protocol was enhanced and designed to be completely compliant with that used in previous studies[10,26] in order to maintain forward compatibility without modifying the previously-reported fitting protocol for monohalogenated molecules. The accuracy, stability, transferability, and physical meanings of the parameters were also discussed in this study. The force field fitting protocol developed was highly efficient, allowing the entire procedure to be parallelized in distributed computing systems.

## 2. Theory and Computational Details

In the PEff model, the halogen bond is a typical noncovalent bond consisting of three fundamental interactions: electrostatic, repulsive/dispersive, and polarization interactions.[10] The PEff model can accurately reproduce the potential energy surface of halogen bonds at the quantum mechanics level. The electrostatic interaction in the PEff model is expressed as follows:

$$V_{\text{elst}}(r_1, R, r) = Q \cdot [\exp(-\alpha r_1 - \beta R) - exp(-\varsigma r)]/r \qquad (1)$$

where $r_1$ is the coordinate in the equatorial area, $R$ is the distance from the halogen atom along the R–X axis direction, and $r$ is the halogen bond length. The anisotropic charge distribution is represented by the combination of a negatively charged sphere and a positively charged ellipsoid, as shown in Figure 1(a). The parameters ($\alpha$, $\beta$, $\zeta$, and $Q$) are derived from the *ab initio* electrostatic potential to mimic realistic charge distributions. Parameters $\alpha$ and $\beta$ are the factors of the equatorial area and along the R–X axis direction, respectively. The relationship $\alpha < \beta$ revealed that the anisotropic charge density on the covalent halogen atom and the electrostatic potential along the R–X axis were more positive than those in the equatorial area. The variable $\zeta$ is a factor of the negatively charged sphere, whose shape is related to $\zeta$ to a certain degree. The factor $Q$ is the fitted charge for the covalent halogen atom in the PEff model, where a larger $Q$ indicates a more positive charge of the halogen atom in average. Therefore, the parameters ($\alpha$, $\beta$, $\zeta$, and $Q$) were sufficient to describe the anisotropic charge distribution of halogen atoms in the PEff model.

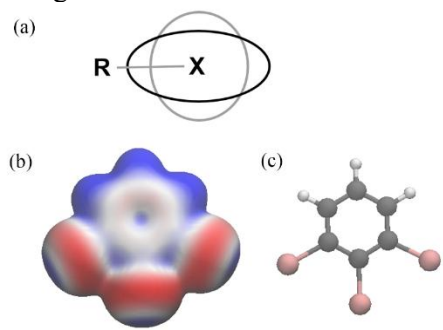

Figure 1. (a) PEff model of a covalent halogen atom; (b) electrostatic potential and (c) structure of 1,2,3-tribromobenzene.

Applying the PEff model for polyhalogenated molecules is a more complex task. As shown in Figure 1(b), 1,2,3-tribromobenzene contains three consecutive covalent halogen atoms. The chemical environment of the first and third bromine atoms should be similar, whereas that of the second bromine atom should be different. However, the 12 electrostatic parameters could not be fitted easily using the least-squares method that was implemented for monohalogenated molecules, because higher degrees of freedom would reduce the likeliness of obtaining the optimal solution in a MaOP.[27,28] Many local solutions can be obtained if the parameter sets are sufficiently large. However, these solutions are usually not optimal, even without any physical significance.

As MaOPs are often transformed into single-objective optimization problems, the optimization of polyhalogenated systems can be approximated into a series of single halogen issues, significantly reducing the difficulty of the optimization. Taking 1,2,3-tribromobenzene as an example (Figure 1(c)), the parameter values can be derived in a stepwise fashion. The first bromine atom can be fitted first using the PEff model, simultaneously treating the other bromine atoms as point charges. When fitting the second bromine atom into the model, the parameters of the first bromine should be fixed. The parameters of the third bromine atom can then be optimized. However, this method causes the fitted parameters to be self-inconsistent because the fitting order of the halogen atoms would lead to non-negligible errors in the results. In addition, the fitted parameters could be extremely large or extremely small. Unstable parameters cannot provide clear physical meanings. Therefore, this fitting protocol requires further consideration.

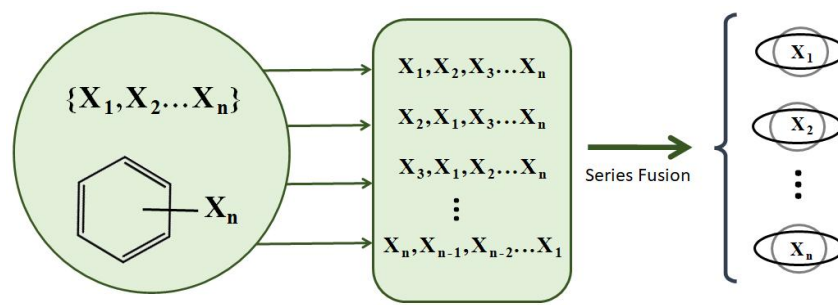

Scheme 1. Parameter optimization protocol for polyhalogenated molecules in the PEff model. Each fitting series (i.e., X1X2…Xn) only provides a partial depiction of the parameters. The series fusion operation can extract the parameters with desirable accuracy, and a simple three-layered feedforward neural network is sufficient.

By traversing all the possible fitting orders of halogen atoms (Scheme 1), the desired parameters were derived via sufficient statistics.[29,30] Each fitting order (i.e., X1X2…Xn) was referred to as a fitting series or a fitting basis. The least-squares algorithm was repeatedly applied to each halogen atom for a specific fitting series to derive a partial depiction of the parameters. The obtained datasets were then fed into the series fusion operation, which could be defined by many machine-learning algorithms, such as a neural network. A simple three-layered feedforward neural network with only one hidden layer was sufficient. Each fitting series could be regarded as the basis for the entire set of the fitting series. This did not mean that the results of each fitting series were incorrect. In contrast, the result of any fitting series is chemically meaningful, as they provide a partial description of the desired parameters. This idea is similar as the role of delocalized resonance structures in the Kekule formula, as discussed in many general chemistry textbooks.

Scheme 1 shows the proposed fitting protocol for parameter optimization. For each fitting series, the halogen atom that appeared before the other atoms was optimized first, fitting as much information as possible. This algorithm, called the greedy algorithm, would make the optimal choice at each step,[31] enabling the entire problem to be solved using a simple three-layered feedforward neural network. This fitting protocol is robust and allows the optimization to be solved from an initial guess of zero. It is also very suitable for parallel computing because map and reduction operations are required for each fitting series. In addition, this protocol can be easily automated on a streaming computing framework such as Hadoop in order to perform high-throughput calculations,[32] even using a laptop.

Finally, the partial atomic charge excluding halogens were calculated using the generalized AMBER force field (GAFF) model.[33] The electrostatic potential of the polyhalogenated molecule was calculated from ab initio calculations at the HF/6-31G* level using the ORCA package.[34] All parameter fitting procedures were implemented using Python scripts. These results were successful, as discussed in the following section.

### 3. Results and Discussions

The parameter fitting protocol proposed in this study is automatic and can be applied to numerous halogenated benzene compounds. The fitted PEff parameters of the halogen atoms were satisfactory. The average errors obtained from the *ab initio* electrostatic potential were of the order of $10^{-6}$, demonstrating the significant improvement in the point charge model of the covalent halogen atoms. For halogen atoms with similar chemical environments, the parameter values would usually fluctuate within a limited range and good parameter transferability would be expected. As the number of covalent bromine atoms increased, a certain chemical significance was observed, as discussed below.

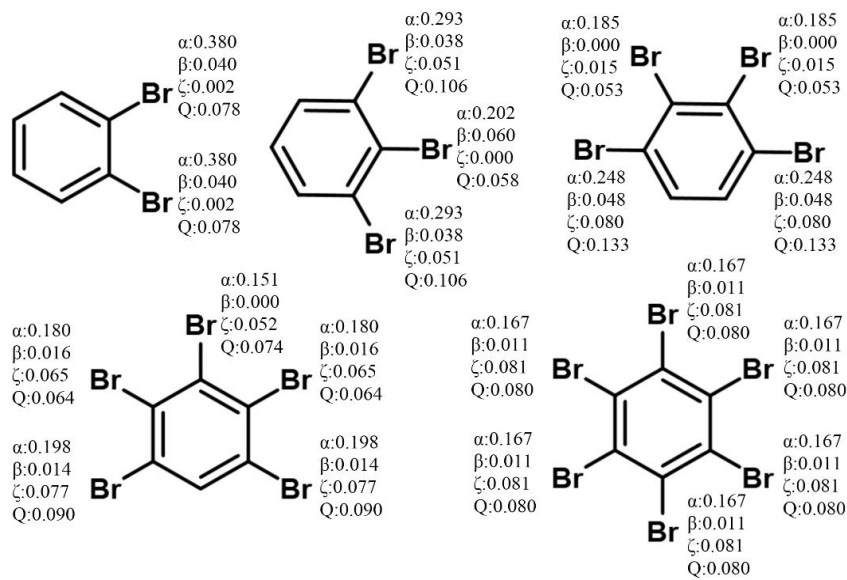

Figure 2. Fitted electrostatic potential parameters ($\alpha$, $\beta$, $\zeta$, and $Q$) of the PEff model for polybrominated benzenes with ortho-substituted covalent bromine atoms.

Figure 2 shows the fitted electrostatic potential parameters ($\alpha$, $\beta$, $\zeta$, and $Q$) for the polybrominated benzenes with ortho-substituted covalent bromine atoms. In 1,2-dibromobenzene (Figure 2), the parameter values for both bromine atoms were very similar due to their comparable chemical environments. The parameter $\beta$ value for dibromobenzene was lower than that of single-substituted bromobenzene,[10] consistent with the variation in the electrostatic potential around the halogen atoms.

The maximum value of molecular electrostatic potential (Vmax) for bromobenzene was 9.68 kcal/mol, whereas that for dibromobenzene was 13.14 kcal/mol for each halogen atom. This suggested that an increase in the number of halogen atoms would lead to an increase in Vmax and a decrease in the ellipsoidal parameters ($\alpha$ and $\beta$).

The fitted parameter sets could rationalize the symmetric features and chemical environments of each halogen atom. For example, two types of bromine atoms exist in the 1,2,3-tribromobenzene molecule, where the chemical environments for the first and third bromine atoms are similar. This was consistent with the interpretation suggested by the fitted parameters, in which different parameter values and electrostatic potential distribution were assigned to the second bromine atom. The Vmax of the first and third bromine atoms was 16.21 kcal/mol, whereas that of the second bromine atom was 15.93 kcal/mol. Two types of bromine atoms could also be present in the 1,2,3,4-tetrabromobenzene molecule. The fitted parameters indicated that the first and fourth bromine atoms were of the same type, while the second and third bromine atoms were of another type. Similar conclusions were obtained for penta- and full-substituted bromobenzene, suggesting that the improved parameter fitting protocol could accurately describe the chemical environment of different halogen atoms without human interventions.

A significant bias could be observed if only one fitting series (such as X1X2…Xn) was applied, or if the series fusion was not applied for parameter fitting. For example, the parameters of the three bromine atoms for tribromobenzene would significantly differ if the optimization algorithm (such as the least-squares method) was directly applied to one possible fitting series. This would lead to a numerical bias greater than 50% for the first and third bromine atoms, which contradicted the similar chemical environments of the two bromine atoms. The series fusion operation could alleviate such unsatisfactory issues, thereby improving the rationality of the PEff model. Note that the single-substituted benzenes could still use the least-squares method, while the polyhalogenated benzenes required a cascade of least-squares optimization protocols. Thus, the series fusion operation yielded a lossless and forward-compatible fitting protocol, providing a good trade-off between accuracy and stability.

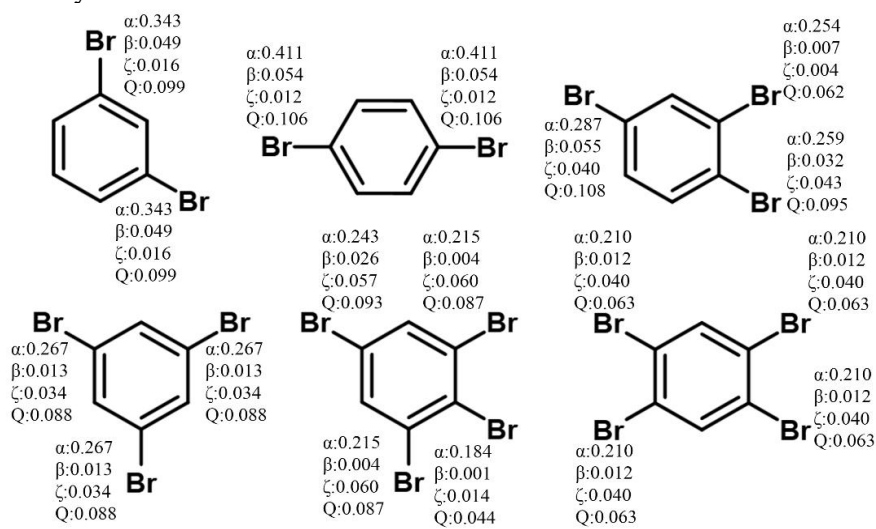

Figure 3. Fitted electrostatic potential parameters ($\alpha$, $\beta$, $\zeta$, and $Q$) of PEff model for the mixed ortho-, meta-, and para-substituted bromobenzene compounds.

Twelve possible polyhalogenated compounds are depicted in Figures 2 and 3. Figure 3 shows the substituent effects of the mixed ortho-, meta-, and para-substituted bromobenzene compounds. The fitted parameters were sensitive to the chemical environments of various bromine atoms. As such, the electrostatic potential along the R–X axis was more positive than that along the equatorial area, as revealed by the relationship $\alpha < \beta$. This resulted in the anisotropic charge density on the halogen atom, demonstrating the σ-hole effect. The symmetry of the parameter values was maintained for the corresponding bromine atoms, such as in dibromobenzene and 1,3,5-tribromobenzene. For 1,2,4-tribromobenzene, the parameter values for each bromine atom varied with their chemical environment. Since the $Q$ value reported in this work was generally positive, the relationship of $\beta < \zeta$ indicated the attractive electrostatic interaction in the halogen bonds.

For the monohalogenated molecules, the exponential parameters (\alpha, \beta, and \zeta) were suggested as an electrostatic interaction index for halogen atoms,14 providing a highly reasonable and meaningful description of the substituent effect for covalent halogen atoms. This PEff model could also be rationalized using a combination of HOMO and LUMO pairs. The HOMO, as the highest occupied molecular orbital, is filled with electrons that carry a negative electron density (negative charge), while the empty LUMO is not filled by electrons, carrying a hole density (positive charge). With the success of frontier orbitals in allowing chemical problems to be understood via the molecular orbital theory, the negatively charged sphere and positively charged ellipsoid can be viewed as further simplification of the HOMO and LUMO in the force field realm. The negatively charged sphere is related to a filled HOMO orbital, and the positively charged ellipsoid is related to an empty LUMO orbital.

While this is a very rough analogy, assumptions can be used to interpret some interesting behaviors of the parameter sets. Taking the 1,2,3-tribromobenzene molecule as an example, the parameter \zeta of the negatively charged sphere was zero, which could be rationalized from the frontier molecular orbitals. The second bromine atom did not participate in the HOMO orbital (Figure 4), which was mainly occupied by the first and third bromine atoms. The electronic density should also be delocalized beyond the second bromine atom. Similar trends were observed for 1,2,3-chlorobenzenes and 1,2,3-iodobenzenes. A smaller \zeta value indicated that the negative charge was more delocalized over a specific halogen atom. These results further verified the rationality of the PEff model for polyhalogenated molecules.

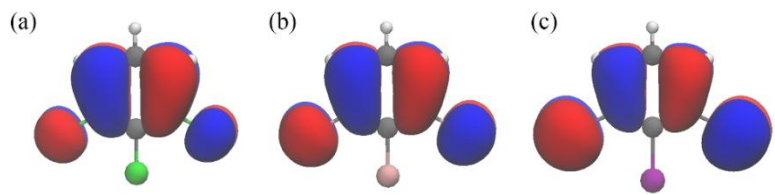

Figure 4. HOMO orbitals of the (a) 1,2,3-trichlorobenzene, (b) 1,2,3-tribromobenzene, and (c) 1,2,3-triiodobenzene. The smaller $\zeta$ value indicated that the negative charge was more delocalized over the specific halogen atom.

## 4. Conclusions

The application of the PEff model on monohalogenated and polyhalogenated molecules requires a sophisticated multi-objective optimization algorithm. This is especially true for the nonlinear parameters in electrostatic interactions, since each covalent halogen atom requires four parameters. To maintain forward compatibility, monohalogenated molecules can use the least-squares method, but polyhalogenated molecules require a cascade of least-squares optimization actions. This fitting protocol was accomplished using a simple series fusion operation. The electrostatic parameters of various ortho-, meta-, and para-substituted benzenes were shown to be mainly affected by the chemical environment around the covalent halogen atoms. The fitted parameter sets were reasonable and stable, exhibiting physical significance. Although the calculation time increased, the fitting protocol was fully parallelizable with a very high computational efficiency. Each fitting was performed on a laptop. The computations also demonstrated possibility to be automatically extended to tens of thousands of halogenated homologues. There are more than one thousand polyhalogenated benzenes with mixed halogen atoms, which would be reported as a dataset in recent studies. I would also prefer to design a statistical molecular library with further molecular simulations from simple systems to complicated ones in the future work.

**Acknowledgements**
This work was financially supported by the Binzhou University (No. 2019Y13) for quantum chemistry calculations on the Aliyun Elastic Compute Service. The author would like to thank Chengbu Liu for his theoretical foundation work of PEff model.

*Science* 2018, *8,* e1327.